\documentclass[10pt]{iopart}

\usepackage{iopams}  
\expandafter\let\csname equation*\endcsname\relax
\expandafter\let\csname endequation*\endcsname\relax
\usepackage{graphicx}
\usepackage{grffile}
\usepackage{amsmath}
\usepackage{float}
\usepackage{cite}
\usepackage{bibentry}
\usepackage[T1]{fontenc}
\usepackage[utf8]{inputenc}
\usepackage{listings}
\DeclareUnicodeCharacter{0301}{\'{e}}
\usepackage{mathptmx}
\begin{document}

\title{Skyrmion Racetrack memory with an antidot}

\author{Aroop Kumar Behera$^{1}$, Chandrasekhar Murapaka$^{2}$, Sougata Mallick$^{1}$, Braj Bhusan Singh$^{1}$, Subhankar Bedanta*$^{1}$}

\address{$^{1}$Laboratory for Nanomagnetism and Magnetic Materials (LNMM), School of Physical Sciences, National Institute of Science Education and Research (NISER), HBNI, Jatni-752050, India}
\address{$^{2}$Department of Materials Science and Metallurgical Engineering, Indian Institute of Technology, Hyderabad, Kandi Sangareddy 502285, Telangana, India}

\ead{sbedanta@niser.ac.in}
\vspace{10pt}
\begin{indented}
\item[]May 2020
\end{indented}

\begin{abstract}
Skyrmion racetrack memory has a lot of potential in future non-volatile solid state devices. By application of current in such devices both spin-orbit torque and spin-transfer torques are proven to be useful to nucleate and propagate skyrmions.  However the current applied during nucleation of successive skyrmions may have unwanted perturbation \emph{viz.} Joule heating and skyrmion Hall effect, on the propagation of previously generated skyrmions. Therefore new methodology is desired to decouple the generation and propagation of skyrmions. Here we present a novel route via micromagnetic simulations for generation of skyrmions from triangular antidot structure in a ferromagnetic nanotrack using local oersted field. Antidots are holes in a magnetic nanoelement.  Multiple skyrmions can be simultaneously generated by incorporating more number of antidots. Controlled skyrmion injection can be achieved by tuning the separation between the antidots that are placed at either end of the nanotrack. Here we propose a novel design to realise skyrmionic racetrcak memory where one can individually generate and manipulate the skyrmions within the nanotrack.
\end{abstract}

\vspace{2pc}
\noindent{\it Keywords}: skyrmion, Dzyaloshinskii–Moriya interaction, magnetic antidot, racetrack memory, spin-transfer torque
%\maketitle
\ioptwocol

\section{Introduction}

Skyrmions~\cite{skyrme1962unified} are topologically protected magnetic states wherein the magnetic moments are arranged in a twisted configuration. This spin configuration has a local minima in magnetic energy, and hence this system tries to remain in the given metastable state when subjected to external perturbations. Due to their topological stability and requirement of relatively low driving current density, skyrmions are potential candidates for future spintronic memory and logic devices~\cite{fert2013skyrmions,nagaosa2013topological,sampaio2013nucleation,buttner2017field,
durrenfeld2017controlled,flovik2017generation,hrabec2017current,lemesh2018current
,yu2017current}. Skyrmions are stabilised in magnetic systems \cite{everschor2018perspective,duong2019stabilizing}, (both bulk and thin films) via an interplay of exchange energy, Dzyaloshinskii-Moriya interaction (DMI) \cite{dzyaloshinsky1958thermodynamic,moriya1960anisotropic}, Zeeman energy, demagnetization energy and magnetic anisotropy \cite{nagaosa2013topological,yu2017current,buttner2015dynamics,woo2016observation}. In recent years skyrmions in ferromagnetic thin films and nanostructures are of great interest due to their potential in skyrmion race track memory and other spintronic applications. Here the system should posses broken inversion symmetry and a proper combination of interfacial Dzyaloshinskii-Moriya interaction (iDMI) and magnetic anisotropy \cite{behera2018size}.\\ 

Utilization of skyrmions for logic and memory devices demands a prior understanding of the efficient methods for their controlled generation and propagation. Recently there has been immense effort devoted to nucleate skyrmions via spin-polarized current, strain induced anisotropy, constrained geometry, etc. \cite{fert2013skyrmions,nagaosa2013topological,sampaio2013nucleation,iwasaki2013current,
jiang2015blowing,liu2017chopping}. Jiang et al. \cite{jiang2015blowing}, have shown the nucleation of skyrmions from chiral stripe domains (CSD) by pushing it through a geometrical constriction under the application of an in-plane current. B{\"u}ttner \cite{buttner2015dynamics} et al. and Everschor-Sitte \cite{everschor2017skyrmion} et al.,  have previously reported nucleation of skyrmions in magnetic track by incorporating local variation in magnetic anisotopy. Further Zhou \cite{zhou2015dynamically} et al., have used local Oersted fields to nucleate dynamically stabilized skyrmions. Although there are several methodologies demonstrated for nucleation of skyrmions, nevertheless it should be noted that  application of a current may affect the stability of the pre-generated skyrmions within the nanotrack. This may lead to issues with device reliability. Therefore a novel route for sequential injection of skyrmions in to a track without interfering with other skyrmions is of utmost importance \cite{fert2017magnetic,finocchio2016magnetic}.

In this paper we report creation of skyrmions from an antidot using micromagnetic simulations by varying the local magnetic field. Magnetic antidots are holes in a thin film or nanostructure which can be easily fabricated via various lithography techniques. It has been shown that magnetic antidots act as pinning and nucleation centers for magnetic domain walls \cite{mallick2015size,mallick2018relaxation}. The edge of the antidots (triangular holes) work as the nucleation centers for the domains due to low anisotropy and high demagnetization energy. In this work we demonstrate that the nucleation of skyrmion from such an antidot strongly depends on a critical ratio of the antidot area with that of the magnetic nanotrack. We further propose a novel route to employ such a system as a skyrmionic race track memory device.

\section{Simulation Details}

We have performed micromagnetic simulations using oxsii solver \cite{donahue1999oommf,rohart2013skyrmion} on a ferromagnetic nanotrack  with antidots at the ends of the sample. The dimensions of the nanotrack are: length $=$ 1000 nm, width $=$ 120 nm, and thickness $=$ 1.1 nm. The antidot is an equilateral triangle with the length of 60 nm. Material parameters used for the simulations are set as the parameters for $Co_{20}Fe_{60}B_{20}$ \cite{jiang2015blowing}. Following are the material parameters used in our simulations: saturation magnetisation $M_s$ $=$ $6.5\times 10^5$ A/m, exchange constant A $=$ $4.5\times10^{-12}$ J/m, perpendicular magnetic anisotropy (PMA) $K$ $=$ $3\times10^5$ J/$m^3$, and the DMI constant $D$ $=$ $1\times10^{-3}$ J/$m^{2}$. The cell size of the simulations was fixed at $2\times2\times1.1  nm^3$. The Gilbert damping constant ($\alpha$) is taken to be 0.008 in all the simulations which is comparable to the value observed experimentally\cite{Braj2018}. The gyromagnetic ratio ($\gamma$) is considered to be 2.211$\times10^5$ m/Amp.s\cite{donahue1999oommf}.

\section{Results and discussion}

The micromagnetic simulations were performed using object oriented micromagnetic framework (OOMMF) package which solves the Landau-Lifshitz-Gilbert (LLG) equation in a ferromagnetic system which is given by:
\begin{equation}
\begin{split}
	\frac{d{\bf m}}{dt}=-|\gamma|\textbf{m}\times \textbf{H}_{eff}+\alpha \bigg( \textbf{m}\times\frac{d\bf m}{dt} \bigg) \\+u \textbf{m}\times\bigg(\textbf{m}\times\frac{\partial \textbf{m}}{\partial x}\bigg)+\beta u  \textbf{m}\times\frac{\partial \textbf{m}}{\partial x}
\end{split}
\end{equation}
 where $\beta$ is the non-adiabatic constant considered to be 0.04. $u$ is the drift velocity of conduction electrons defined as $u=JPg\mu B/(2eM_s)$ where $P$ is the spin polarization. The effective magnetic field of the system $(\textbf{H}_{eff})$ is given by:

\begin{equation}
	\textbf{H}_{eff} = -1/(\mu_0M_s)(\partial E_{total}/\partial{\bf m})
\end{equation} where $\mu_0$ is the permeability in vacuum and $E_{Total}$ is the magnetic energy of the system which consists of contributing energy terms such as exchange ($E_{Ex}$), anisotropy ($E_{Ani}$), DMI ($E_{DMI}$), demagnetization ($E_{Demag}$) and Zeeman energy ($E_{Zeeman}$). We have used the classical model of localized magnetic spins in our system. The total energy of the system can be written as:
\begin{equation}
	E_{Total}=E_{Ex}+E_{DMI}+E_{Ani}+E_{Demag}+E_{Zeeman}
\end{equation}
where, $E_{ex}=-J_{i,j}({\bf m_i}\cdot{\bf m_j})$, $E_{ani}= -K({\bf m_i}\cdot{\bf z})^2$ and $E_{DMI}={\bf{D_{ij}}}\cdot({\bf{m_i}}\times{\bf{m_j}})$, $E_{Demag} = -\mu_{0}/2\int{\bf M\cdot H_d}{\bf dV}$ and $E_{Zeeman}$= -$\mu_{0}{\bf{H}}\cdot{\bf{m_i}}$, respectively. The ${\bf{m_i}}$ and ${\bf{m_j}}$ are magnetization in neighboring unit cells, with each unit cell containing a collections of atomic spins. The $J_{i,j}$ is the Heisenberg exchange coefficient, $K$ is the anisotropy constant, ${\bf{D_{ij}}}$ is the interfacial DMI interaction strength, ${\bf{H_{d}}}$ is demagnetization field  and $\mu_0{\bf{H}}$  is the external applied magnetic field. It is noteworthy to mention that micromagnetic simulations make use of the continuum hypothesis where an average magnetization is used for calculation in a cell where it is assumed that the cell has a collection of spins and the variation of the magnetization is continous. 

The propagation of skyrmion in a magnetic nanotrack is described by the Thiele equation \cite{thiele1973steady} which is given by:
\begin{equation}
{\bf{G}}\times{\bf{v}}-\alpha{D}{\bf{v}}+{\bf{F}} = 0
\end{equation}
where ${\bf{G}}$ is the gyromagnetic coupling vector given by ${\bf{G}}=G {\bf{j}}$. Here $G$ is given by $G=(4\pi Q)$ where $Q$ is the topological charge, and ${\bf{j}}$ is the unit vector along $z$ axis. ${\bf{v}}$ is the velocity of the spin texture (skyrmion in our case). The first term in the equation explains the magnus force acting upon the skyrmion. ${D}$ represents the dissipative force tensor where the $xx$ and $yy$ components have the same value as ${D}$, and the $zz$ and the rest $antisymmetric$ terms are $zero$. ${\bf{F}}$ represents the force acted upon the skyrmion from the surrounding environment.

\begin{figure}[hb!]
	\centering
	\includegraphics[scale=0.06]{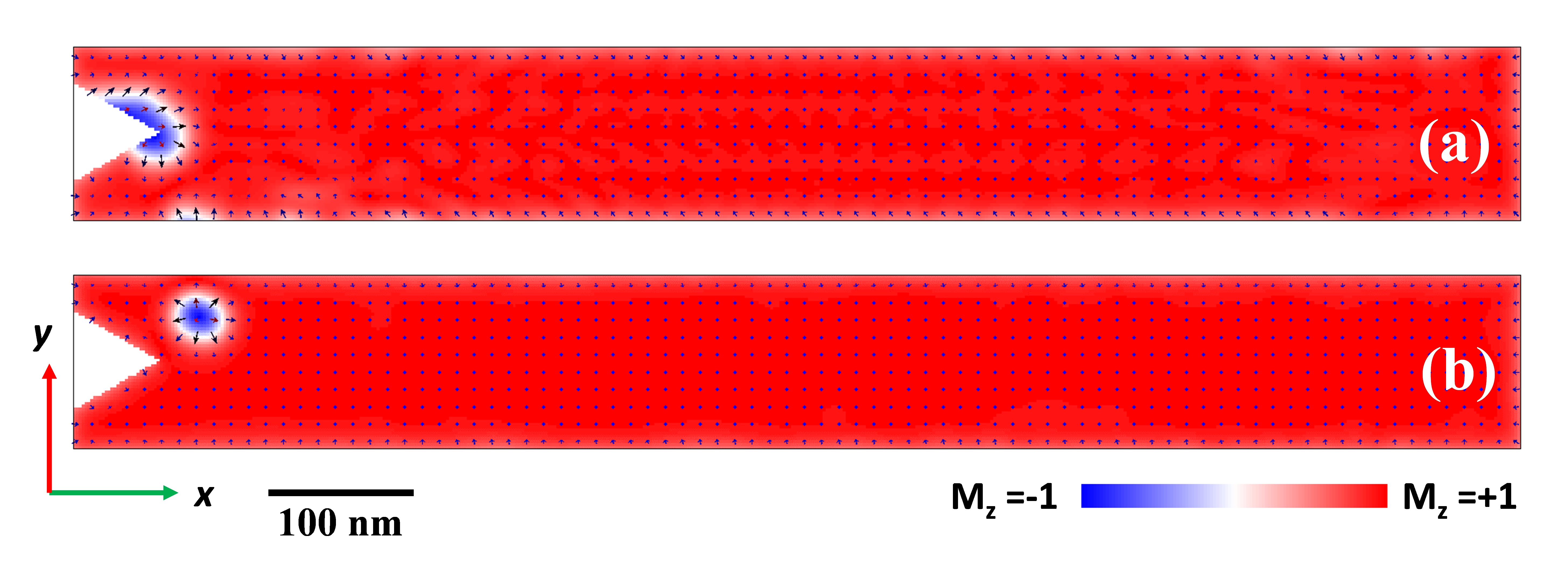}
	\caption{(a) Nucleation of domain upon reducing the field from $70$mT to $35$mT for 0.4 ns. (b) Depinning of the domain from the antidot and the evolution of the skyrmion after 2 ns. The green and red arrows denote the sample map in the $xy$ plane. The $z$ direction is normal to the plane of observation. The color scale and the scale bars are indicated in the figure. } 
	\label{fig:Figure_1}
\end{figure}
\begin{figure}
	\centering
	\includegraphics[scale=0.12]{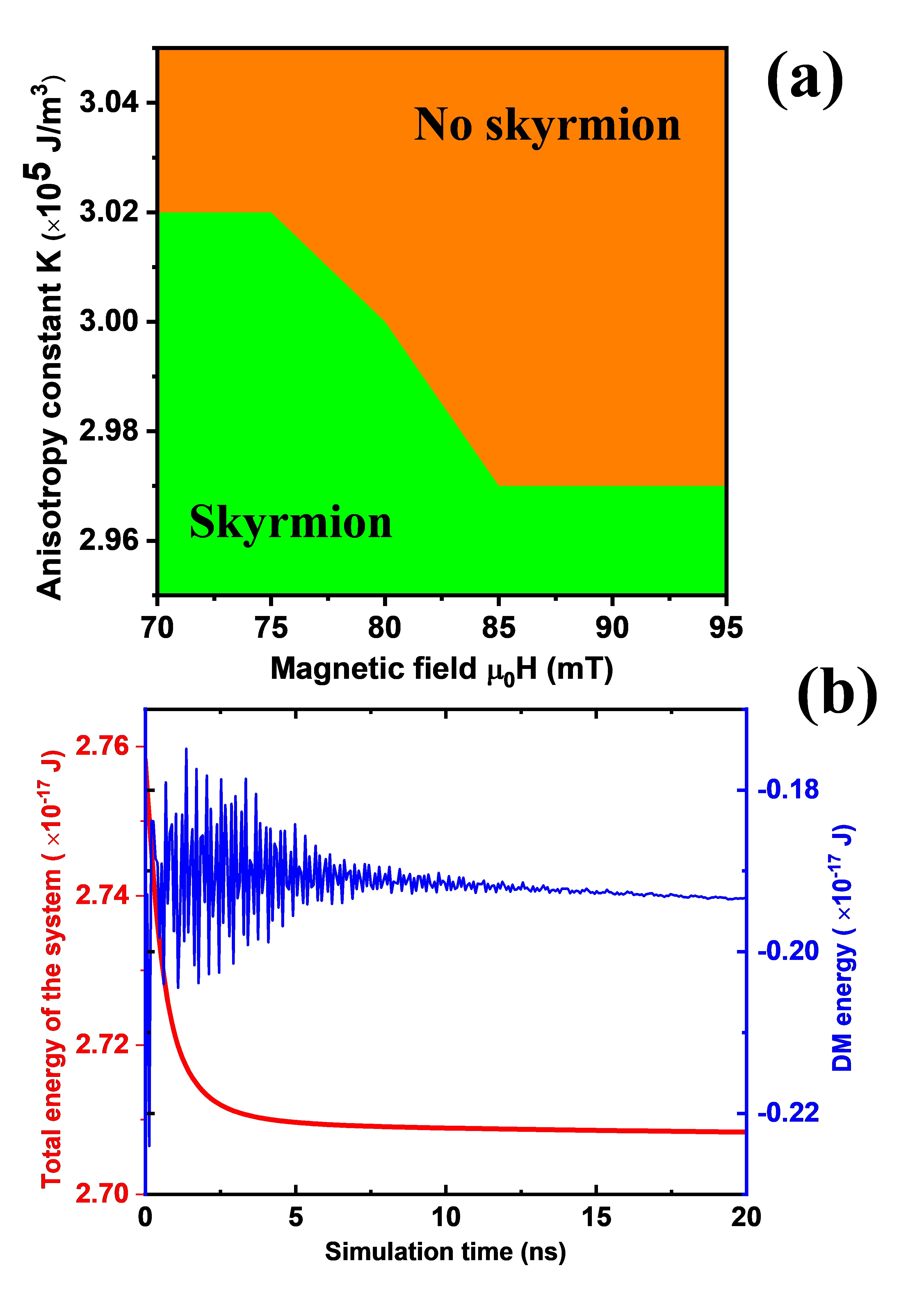}
	\caption{(a) Phase plot for the skyrmion formation as a function of the perpendicular magnetic anisotropy ($K$) and applied magnetic field ($\mu_{0}H$) at a constant $D$ = 1.0 mJ/m$^2$. (b) The total (red) and DM  (blue) energies vs time during the evolution of the skyrmion shown in Figure 1.}
	\label{fig:Figure_2}
\end{figure}

In order to generate skyrmions in the nanotrack, the sample was saturated along +z direction under the application of an external magnetic field of 70 mT. Further, the field was reduced to 35 mT to nucleate a domain with opposite magnetization at the vicinity of the antidot. Subsequently, the magnetic field was increased back to 70 mT to depin the domain from the antidot vertex. Figure 1 shows the snapshot of the magnetization state during the evolution of skyrmion as a function of time. When the field was reduced to 35 mT the domain evolution occurs which is shown in Figure 1(a). The subsequent depinning of the nucleated domain from the antidot vertex occurs and after 2 ns it leads to the formation of an isolated skyrmion  as shown in Figure 1 (b). The skyrmion is stabilized due to the presence of DMI in the system. It should be noted that the magnetic field is needed to be kept at 35 mT for a very short interval of time (a few hundreds of ps) for the nucleation of domain with opposite magnetization which will eventually stabilize into a skyrmion. 
If the field value is less than 35 mT then we have obsered domain nucleation from all the edges of the nanotrack (see Fig. S1 in the supplementary information). However for 35 mT a controlled nucleation of skyrmion occurs at the antidot. Similarly the pulse duration is also quite important. Further if the field value is more than 40 mT then no skyrmion is stabilized.  Fig. S2 in the supplementary information shows that if the pulse duration is 0.2 ns or lower then no skyrmion generation occurs, however, for 0.25 ns a skyrmion is generated at the antidot. These results show that for a set of $K$, $D$ and $H$, there are critical values for the field and pulse duration at which the skyrmion generation can occur.

\begin{figure}
	\centering
	\includegraphics[scale=0.065]{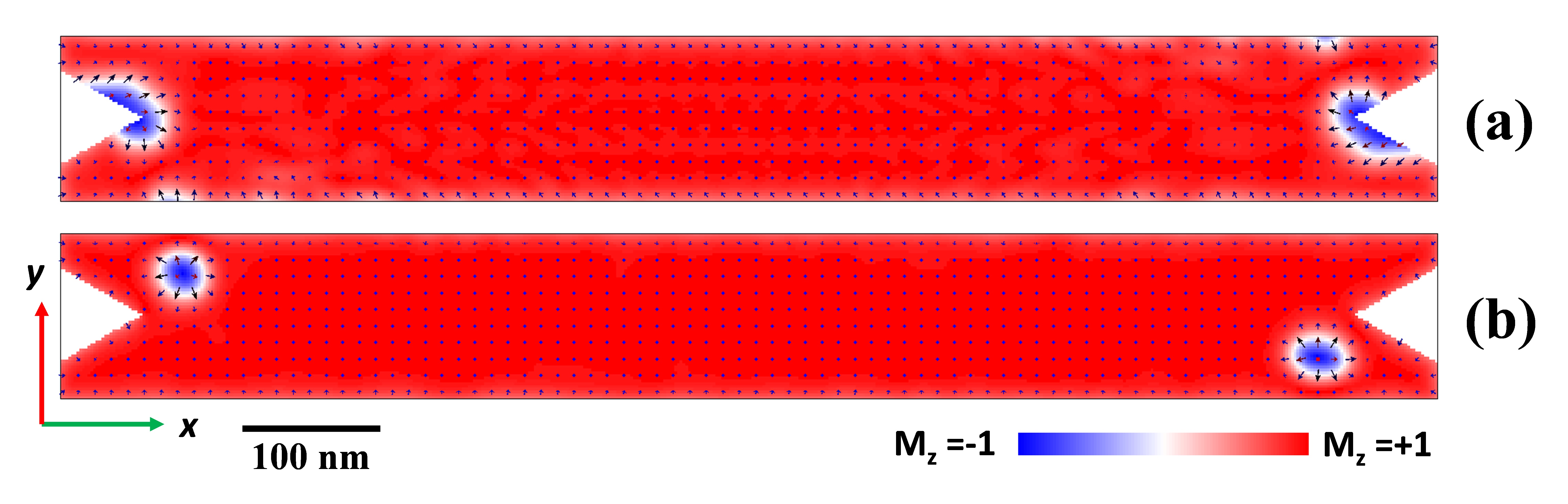}
	\caption{(a) and (b) show the evolutionary steps of creation of two skyrmions from two triangular antidots placed at opposite ends of a magnetic nanotrack at applied fields of 35 and 70 mT, respectively.}
	\label{fig:Figure_3}
\end{figure}

To gain an insight into the dependence of shape of the antidot on skyrmion nucleation, we have performed several simulations by considering square and semi-circular shaped antidots. The results corresponding to the semi-circular and square shaped antidots are presented in supplementary information S3. We have observed that either the field required for reversal maybe higher to nucleate the domain or the time required for the reversed domain to form skyrmion is longer in these shapes. Hence we note that the notch in the shape of a triangular antidot is an efficient way for the formation of the skyrmions in such a system. 

\begin{figure}[ht]
	\centering
	\includegraphics[scale=0.09]{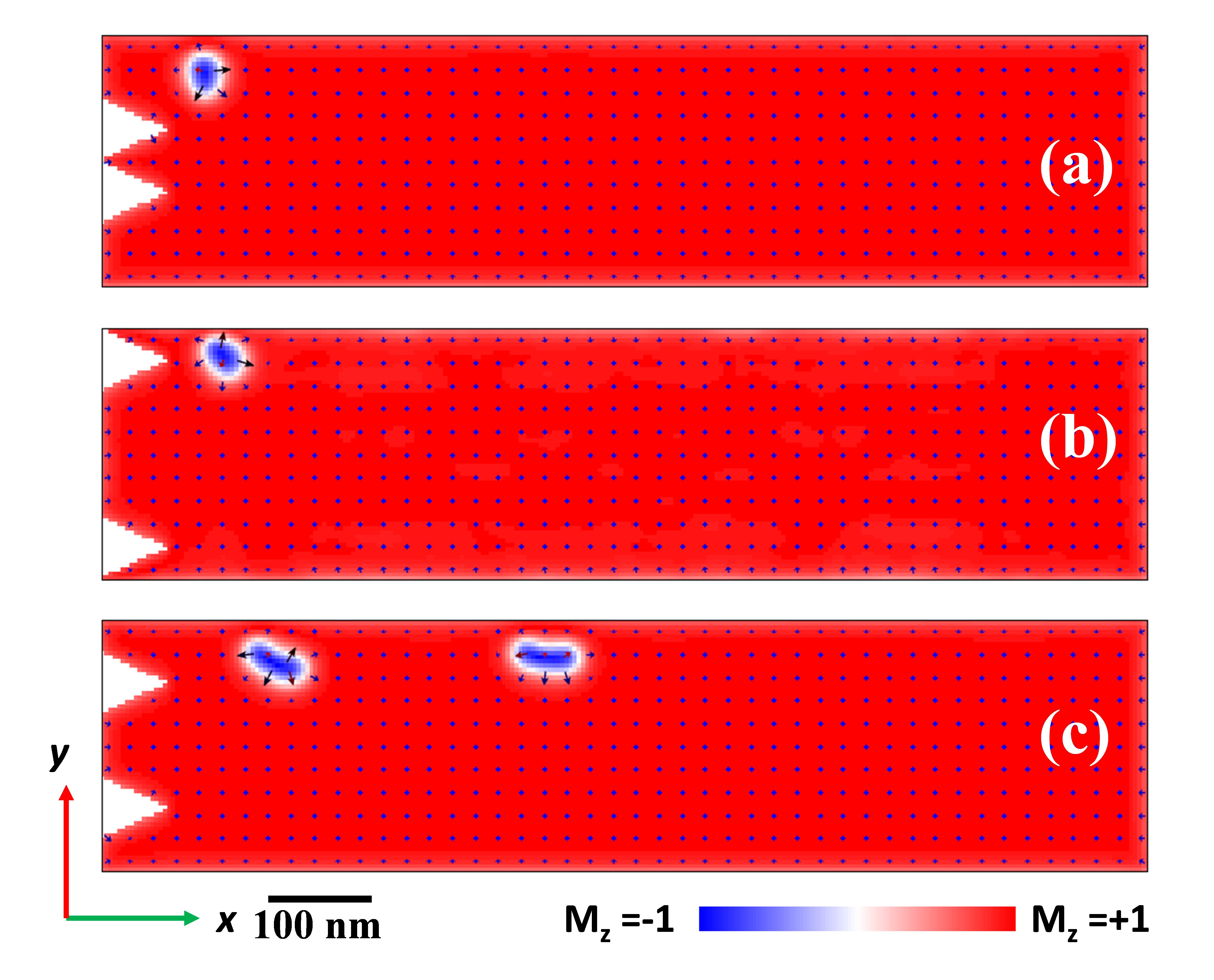}
	\caption{Skyrmion evolution from two triangular antidots placed one above the other and separated by different edge-to-edge distances such as (a) $d$ = 0 nm, (b) $d$ = 120 nm, and (c) $d$ = 60 nm.}
	\label{fig:Figure_4}
\end{figure}

The effect of magnetic field strength as well as the PMA on the skyrmion nucleation is presented in figure 2 (a). It should be noted that skyrmion nucleation occurs for certain combination of magnetic field and $K$ which is presented in the phase plot shown in figure 2(a). For example, at $K$ = $3 \times 10^{5} J/m^3$, it was observed that skyrmion nucleation is possible at relatively lower fields and when the field was increased beyond 80 mT no skyrmion was obtained. At higher field values the nucleated domain subsequently gets annihilated into the antidot. We have also observed that the skyrmion size varies as $\sqrt{D/B}$ as reported in literature \cite{you2015critical}. Figure 2(b) shows the time evolution of the total energy and DM energy for the simulation results as shown in Figure 1. The evolution of the skyrmion from the nucleated domain can be understood from the oscillatory behavior of the DM energy for the first few nanoseconds. While the DM energy is stabilizing the total energy of the system approaches minimum equilibrium.  

The above results show a clear methodology for the nucleation of skyrmions from an antidot. The demagnetization energy helps in initiating the magnetization reversal at the antidot edge by nucleating a domain wall. This nucleated domain further evolves into a skyrmion due to the competition among various energies such as $E_{Ex}$, $E_{Ani}$, $E_{DMI}$, $E_{Demag}$ and $E_{Zeeman}$.  If we apply the field pulse less than the critical time then the nucleated domain does not evolve into the skyrmion rather it gets annihilated. It should be noted that the critical time may vary depending on the relative strength of the energies present in the system as shown in Supplementary figure S2.

In order to create multiple skyrmions in a single nanotrack, we considered two antidots of same size at the two opposite ends of the nanotrack. As expected, in the triangular antidot case, two different skyrmions originated from the opposite ends of the track at a magnetic field of 35 mT, as shown in Figure 3(a). Further when the field value was increased to 70 mT the skyrmion was evolved fully which is shown in Figure 3(b).

The movement of the skyrmion generated on either ends of the nanotrack can be explained from equation (4). Upon solving the equation, it is evident that the deviation of skyrmion path in the nanotrack is given by: $\theta=tan^{-1}\frac{v_y}{v_x}=\frac{G}{\alpha D}$ which is independent of the term $F$ in equation(4). Since the velocity component of skyrmion in the $x$ direction is opposite in nature at the ends of the track, the angular deflection $\theta$ is opposite on the both ends of the track after they move significantly away from the confining forces acting on them from the antidots.

Another set of two triangular antidots has been introduced to develop two skyrmions from one end of the nanotrack. In this configuration, two antidots with similar dimension were placed on the left side of the nanotrack by placing one below the other. It was observed that there was no nucleation of any domain from the antidots when the width of the nanotrack was identical to the shape shown in figure 1. In order to generate a domain from the antidot, the width of the nanotrack had to be doubled. Three different antidot configurations were studied where the antidots were placed at a certain spacing ($d$) between them such as (a) close to each other (i.e. $d$ = 0 nm), (b) far away from each other (i.e. $d$ = 120 nm) and (c) equidistant from the center line (x-axis) of the nanotrack (i.e. $d$ = 60 nm). It must be noted that domain nucleation was possible at a smaller field of 30 mT as compared to the previous case of 35 mT for a single antidot case because of the change in the shape induced anisotropy due to change in dimension. Further it was observed that only one skyrmion was nucleated for cases (a) and (c) and for case (b) formation of two different skyrmions was observed. The results for these three cases are shown in Figure 4 (a) \textendash (c). 
	
\begin{figure}
	\centering
	\includegraphics[scale=0.045]{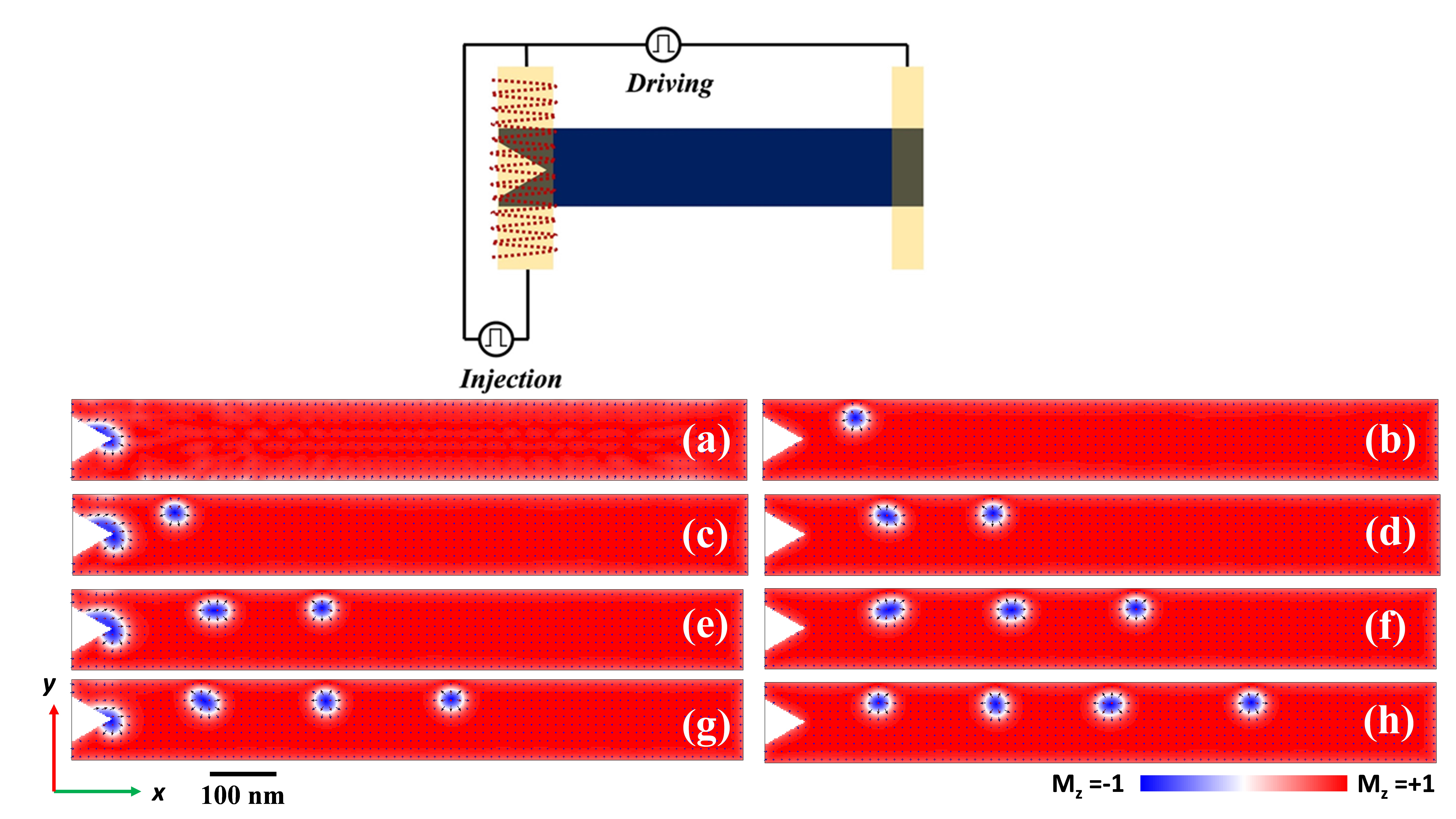}
	\caption{Schematic design for skyrmionic race-track memory device. (a), (c), (e), and (g) represent nucleation of a skyrmion from the antidot edge under the application of Oersted field generated by current pulse. (b), (d), (f), and (h) represent propagation of the nucleated skyrmion through nanotrack under spin-transfer torque.}
	\label{fig:Figure_5}
\end{figure}

The interaction of the two evolving skyrmions nucleating from the antidots can be nullified by increasing the magnetic sample area in the nanotrack. The increase in the magnetic area near the antidots, facilitates the free growth of the two evolving skyrmions without the influence of one on the other. So, the evolving magnetic structure nucleated from both the antidots can now freely develop into two independent skyrmions. It must be noted that, because of the large sample area, the magnetic sample boundary is quite far from the nucleation point of the skyrmions. Therefore the skyrmions do not experience any confining forces from the boundary and/or from another neighboring skyrmion. One can further observe the relatively high transverse motion of skyrmions because of the skyrmion Hall effect. So, the generated skyrmions from the antidots first start moving towards the sample edges. By applying current the skyrmions can be moved in the nanotrack via spin-transfer torque. Hence it is understood that multiple skyrmions can be formed in the nanotrack by increasing the number of antidots until and unless, one evolving skyrmion does not interfere in the development of another during the process of nucleation.

In figure 5, we propose a novel design to demonstrate the skyrmionic racetrack memory. The schematic design presented in figure 5 is conceptually similar to the existing prototype of domain wall based racetrack memory \cite{parkin2008magnetic}. It should be noted that, the simulations presented so far in this work consider a global external magnetic field. However, here we show that  one can apply spatially varying local magnetic field to generate skyrmions. In this design, we have employed a metallic strip line which is transverse to the nanotrack that is overlapped with the antidot (kept at the left hand side of the nanotrack). When current is flown through this metallic strip, Oersted field can be generated to switch the magnetization of the underlying nanotrack. In this way, one can nucleate skyrmions without the application of any external global magnetic field. Once the skyrmion starts evolving (Figure 5 (a)), the current pulse is removed from the transverse metallic strip so that the same skyrmion is stabilized under the influence of DMI as shown in figure 5(b). Then a current pulse (current density J = $5 \times 10^{11}$ A/m$^2$ and the pulse duration = 4 ns) is applied through a nanotrack along x-axis that drives the skyrmion towards the right edge. Again, by flowing current through the metallic strip (along the y-axis) one can locally generate a second skyrmion without affecting the previously nucleated skyrmion (figure 5(c)). This process can be repeated and in this way we can simultaneously create and drive multiple skyrmions in the same nanotrack (figure 5 (d)-(h)). Using local Oersted field to generate skyrmions has several potential advantages compared to the conventional in-plane current induced nucleation. Because in-plane current induced nucleation, skyrmions that were generated earlier in the nanotrack may also get affected as critical current density for skyrmion motion is relatively low. 

\section{Conclusion}

In summary, we present a novel route to nucleate skyrmions from the edge of the triangular antidot inside a nanotrack. The nucleation of the skyrmions strongly depends on shape of the antidot. We further show that multiple skyrmions can be generated by using multiple antidots carefully separated from each other. Most importantly we propose a novel design of skyrmion based racetrack memory device. Here, the skyrmions can be simulatenously generated and driven by using spatially varying local magnetic fields. We believe that our work will present a guidemap towards future skyrmionic based spintronic devices. Further the effect of shape (e.g. isosceles) and size of the triangular antidot on the nucleation of skyrmions will be investigated in a future work.

\section{Acknowledgments}
We thank Department of Atomic Energy (DAE) and Info-French collaborative project supported by CEFIPRA, Govt. of India for providing the funding to carry out the research work. CM would like to thank Department of Science and Technology (DST), Govt. of India for funding under Early Career Research Award (ECR/2018/002664). BBS acknowledges DST, Govt. of India for INSPIRE Faculty Fellowship.

\medskip
\bibliographystyle{iopart-num}
\bibliography{References}

\end{document}